\documentclass[11pt]{iopart}

\pdfoutput=1

\usepackage{graphicx}
\usepackage{amssymb}
\usepackage{color}

\def \f{\frac}

\newcommand{\xpct}[1]{\langle{#1}\rangle}    

\newcommand{\beq}{\begin{equation}}        
\newcommand{\eeq}{\end{equation}}
\newcommand{\bea}{\begin{eqnarray}}
\newcommand{\eea}{\end{eqnarray}}
\newcommand{\non}{\nonumber}

\begin{document}

\title{Real-space structure of the impurity screening cloud in the Resonant Level Model}

\author{Shreyoshi Ghosh$^1$, Pedro Ribeiro$^{1,2}$, and Masudul Haque$^1$}

\address{$^1$
 Max-Planck-Institut f\"{u}r Physik komplexer Systeme,
  N\"{o}thnitzer Stra{\ss}e 38, D-01187 Dresden, Germany}

\address{$^2$ Centro de F\'\i sica das Interac\c c\~oes Fundamentais,
Instituto Superior T\'ecnico, Universidade de Lisboa,
Av. Rovisco Pais, 1049-001 Lisboa, Portugal}

\begin{abstract}

We present a detailed investigation of the impurity screening cloud in the Resonant Level Model.
The screening is visible in the structure of impurity-bath correlators as a function of distance
from the impurity.  We characterize the screening cloud through scaling analyses of impurity-bath
correlators and entanglement entropies.  We devise and study several situations where the screening
cloud is destroyed or modified: finite temperatures, energetic detuning of the impurity from the
chemical potential, and the situation of an unconventional bath with diverging density of states.

\end{abstract}

\maketitle

\section{Introduction} 

The concept of a ``Kondo screening cloud'' has caused much discussion and controversy in the field
of quantum impurity physics.  Single-impurity models generally possess an emergent energy scale.
The most famous is perhaps the celebrated Kondo temperature for the single-impurity Kondo model,
\cite{kondo, hewson} but similar energy scales appear in the single-impurity Anderson model
\cite{hewson, Anderson_model_PRB61} and the interacting resonant level model \cite{IRLM_original,
  IRLM_recent}.  There is a length scale $\xi$ corresponding to this energy scale, which suggests
that the bath surrounding the impurity is affected differently at distances less than $\xi$ from the
impurity than at larger distances $x>\xi$, i.e., there should be a screening cloud of radius $\xi$
surrounding the impurity.

Although the impurity screening cloud is difficult to observe directly experimentally, calculations
have shown that this impurity lengthscale does in fact appear in real-space properties of the bath.
The properties (e.g., persistent current or conductivity) of a mesoscopic device containing a Kondo
or Anderson impurity has been found to behave differently if the device size is larger or smaller
than the size of the Kondo cloud \cite{AffleckSimon_PRL01, SimonAffleck_PRB01, SimonAffleck_PRL02, SimonAffleck_PRB03,
  HandKrohaMonien_PRL06}.  Numerical and variational calculations have found
real-space properties (e.g., impurity-bath correlation functions, distortion of local density of
states, entanglement properties, etc) to be different for $x<\xi$ and $x>\xi$
\cite{BusserAndaDagotto_PRB10, Simonin_arXiv07, Goth_Assaad, 
  Holzner-etal_PRB09, Borda, BordaGarstKroha_PRB09, Saleur, MitchellBulla_PRB11,   Laflorencie, SougatoBose_kondocloud}, for Anderson and
Kondo models and for spin-chain versions of the Kondo model.
These recent results support earlier perturbative calculations of real-space structure \cite{BarzykinAffleck_PRB98, ishii,
  SorensenAffleck96}.

In this work, we focus on the screening cloud around the impurity in the resonant level model (RLM):   
\begin{equation} 
H_{\rm RLM}=  \sum_k \epsilon_k c_k^{\dagger}c_k  - \frac{J'}{\sqrt{\mathcal{L}}} \sum_k (d^{\dagger}c_k
+c_k^{\dagger}d) ~+~ \epsilon_d d^{\dagger}d .
\label{eq_RLM}
\end{equation}
where $c_k$, $c_k^{\dagger}$ are the bath fermion operators at momentum $k$ and $d$, $d^{\dagger}$
are the fermion operators at the impurity site, $\epsilon_k$ is the dispersion of the bath fermions,
$J'$ is the hopping strength between impurity and position $x=0$ of the bath, $\mathcal{L}$ is the
bath size, and the on-site potential $\epsilon_d$ is generally tuned to the bath chemical potential.
Here $x$ represents the distance from the impurity.  Our results are mainly for one-dimensional (1D)
baths, but much of the discussion is expected to be valid for any dimensionality.  For small $J'$,
the RLM possesses a small energy scale and correspondingly a large length scale, depending as
$(J')^{-2}$ on the coupling.  In this article we analyze the real-space structure appearing at such
length scales.

The resonant level model appears as solvable limits of the interacting resonant level model (IRLM)
\cite{IRLM_original}, the single-impurity Anderson model (SIAM) \cite{Anderson_model_PRB61, hewson},
and the anisotropic Kondo model \cite{hewson, toulouse}.  Each of these impurity models have a known
energy scale and associated length scale.
The IRLM, $H_{IRLM} = H_{RLM}+Vd^{\dagger}dc_{x=0}^{\dagger}c_{x=0}$, has an impurity-bath
interaction $V$.  The length scale is known to depend on $J'$ as a power law, $J'^{-\alpha(V)}$, with
the interaction-dependent exponent $\alpha(V)$ taking the value $\alpha(0)=2$ at the RLM point
\cite{IRLM_original, IRLM_recent}.  
The SIAM contains two copies (spins $\uparrow$ and $\downarrow$) of the RLM, with an on-site
interaction $U$ between the two spin species.  The SIAM has an ermergent energy/length scale
\cite{Anderson_model_scale} and the appearance of this scale in the spatial dependence of
correlation functions have been explored in Ref.~\cite{Holzner-etal_PRB09}.  

For the isotropic Kondo model, the energy scale is the Kondo temperature, given by the well-known
expression $T_K=D\exp\left[-1/\rho(\varepsilon_F)J_K\right]$.  (Here $J_K$ is the Kondo coupling,
$\rho(\varepsilon_F)$ is density of states of the conduction electrons at fermi energy and $D$ is
the band width.)  The spatial behavior of the impurity-bath spin-spin correlator has been explored
earlier in Refs.\ \cite{ishii, BarzykinAffleck_PRB98, SorensenAffleck96} and more recently in
Refs.\ \cite{Borda,BordaGarstKroha_PRB09}.
The expression for the energy scale is considerably more complicated for anisotropic Kondo
couplings, but becomes simpler at a special value of the anisotropy called the Toulouse point
\cite{hewson, toulouse}.   At the Toulouse point, the Kondo model can be mapped to the non-interacting
RLM.  Because of solvability, the Toulouse point is widely used in many studies of Kondo physics.
For example, it has been used for non-equilibrium calculations for Kondo
impurities \cite{Toulousepoint_current_noneq, kehrein, medvedyeva_kehrein_arxiv13,
  VasseurHaasSaleur_PRL13}. Spatial structures in the bath have been studied for the Toulouse point
in Ref.\ \cite{trautzettel_PRL13} for helical edge states serving as baths, and in
Ref.\ \cite{medvedyeva_kehrein_arxiv13} in the context of time evolution.

Our work belongs to this general theme of exploring emergent length scales in impurity models
through the study of real-space profiles.  We concentrate primarily on the spatial dependence of the
two-point impurity-bath correlator $\xpct{d^{\dagger}c_x}$, i.e., the equal-time Greens function or
the one-body density matrix.  This is the natural analog, for the RLM, of the impurity-bath
spin-spin correlator $\xpct{\vec{S}_{\mathrm{imp}}\cdot\vec{s}_x}$ commonly used in studies of the
screening cloud in the Kondo and Anderson models \cite{Borda, Holzner-etal_PRB09,
  BordaGarstKroha_PRB09, Goth_Assaad}. 

We present analytic expressions for the model as written in Eq.\ \ref{eq_RLM}, and numerical
calculations for specific lattice implementations.  The lattices are described in Section
\ref{sec_geometries}.  We find the structure of the screening cloud to be very similar for various
lattice geometries, in contrast to some other impurity situations (spin chains, persistent currents
through rings) where the physics can depend markedly on impurity geometry
\cite{EggertAffleck_PRB92, SimonAffleck_PRB01}. 

The correlator $\xpct{d^{\dagger}c_x}$ has oscillations with period equal to the Fermi wavevector
$k_F$ . As in Ref.\ \cite{Borda}, the structure of the screening cloud is seen by analyzing
the envelope of these oscillations.
The analytic expressions for $\xpct{d^{\dagger}c_x}$ presented in Section \ref{sec_greens_function}
show clearly different behaviors for $x<\xi$ and $x>\xi$; the envelope depends logarithmically on
distance within the screening cloud and shows the expected Fermi liquid behavior ${\sim}x^{-1}$ at
larger distances.  The width of the spectral function scales as $J'^2$; we identify this
energy/temperature as the RLM analog of the Kondo temperature.  From these results, with relatively mild
assumptions, one can predict also aspects of the structure of the Kondo cloud in the SIAM (Section
\ref{sec_Anderson_comments}).  
%

In numerical calculations on finite lattices (Section \ref{sec_finite_size}), boundary effects
modify the cloud shape in geometry-dependent ways.  To see the crossover from $-\ln{x}$ to $1/x$
behavior clearly at reasonable system sizes, we have combined multiple values of the coupling $J'$.
The data sets show good scaling collapse.

In Section \ref{sec_detuning} we show what happens to the screening cloud when the impurity level
energy ($\epsilon_d$) is detuned away from the chemical potential.  The detuning induces an
intermediate region in the spatial profile of $\xpct{d^{\dagger}c_x}$, which gradually encroaches
toward smaller distances with increasing detuning and destroys the $-\ln{x}$ behavior within the
Kondo cloud.

In Section \ref{sec_entanglement} we characterize the screening cloud using the quantum entanglement
of a region of size $\ell$ containing the impurity with the rest of the bath.  Unlike
$\xpct{d^{\dagger}c_x}$, we do not have analytic predictions for the size dependence of the block
entanglement entropy, but the numerically determined entanglement shows clear $\ell/\xi$ scaling.

In Section \ref{sec_next_nearest_neighbor} we engineer an unconventional bath that has a divergent
density of states at the Fermi energy, due to part of the dispersion being not linear but quadratic
at the Fermi surface.  The resulting screening involves an additional scale, and the long-distance
power-law decay of $\xpct{d^{\dagger}c_x}$ now has a ``non-Fermi-liquid'' exponent.

\section{\label{sec_geometries}  Lattice geometries: embedded, external, and endpoint impurities}

\begin{figure}
\centering \includegraphics[width=0.6\linewidth]{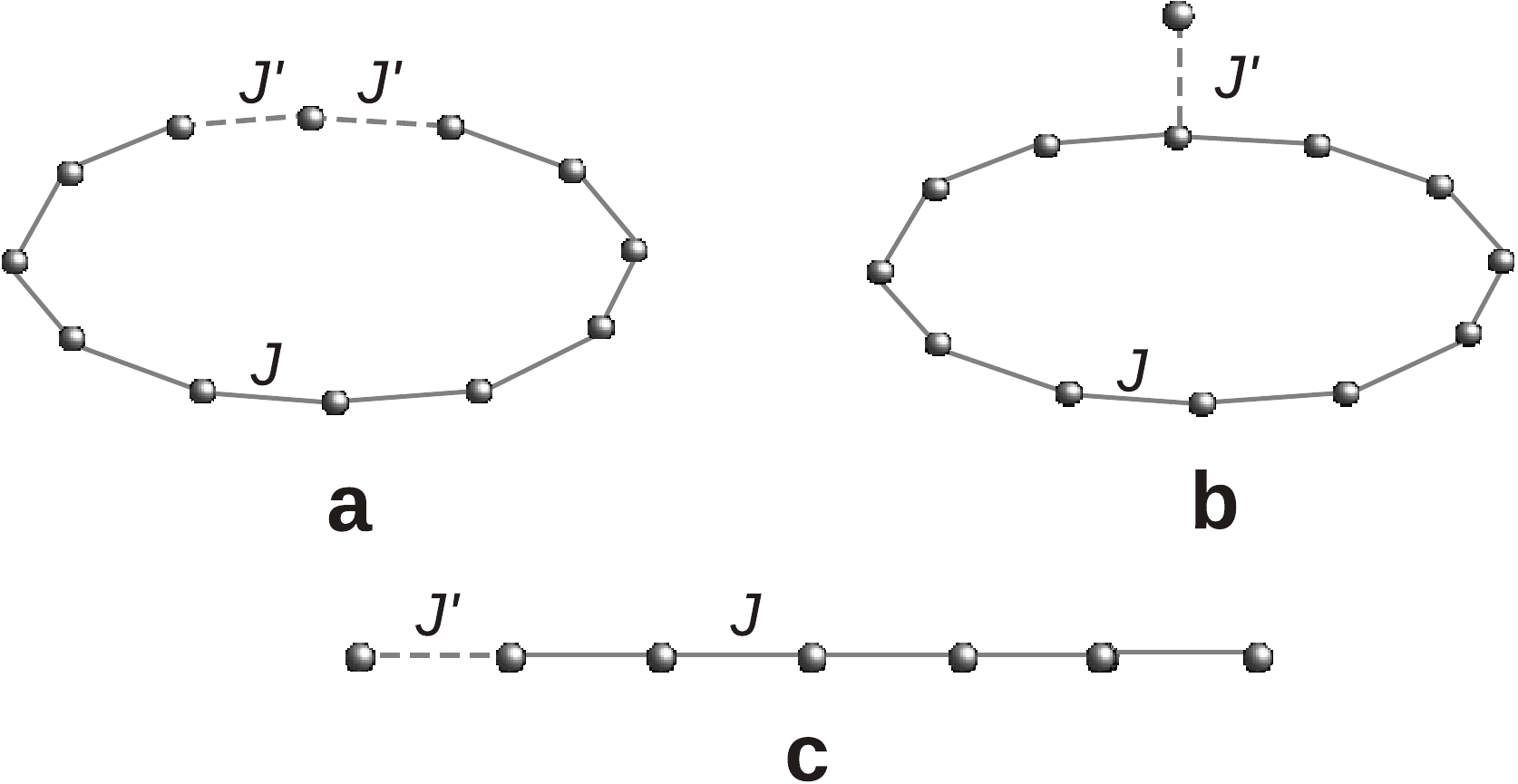}
\caption{Geometries used for lattice realizations of the RLM.  The impurity site
   can be  (\textbf{a})
  Embedded, (\textbf{b}) extrenal, or (\textbf{c}) end-coupled.}
\label{lattice geometries}
\end{figure}

The RLM, given in Eq.~(\ref{eq_RLM}), describes the resonance of an impurity level tunnel-coupled to
a bath of spinless fermions (``conduction electrons'').  We will use 1D tight-binding lattices of
non-interacting fermions to realize the conduction bath.  The impurity level at site $i_{imp}$ is
coupled to this bath with a hopping strength $J'$ much weaker than the hopping $J$ within the bath.
As shown in Figure \ref{lattice geometries}, we consider three different geometries (locations of
$i_{imp}$): (\textbf{a}) embedded, (\textbf{b}) external, and (\textbf{c}) end-coupled RLMs. The
Hamiltonians describing these three geometries are:
\begin{equation} 
H_{emb} = -J\left[ \sum_{i=-L}^{-2} + \sum_{i=1}^L \right]
(c_i^{\dagger}c_{i+1}+ \mathrm{h.c.}) 
- J' \left[  (c_1^{\dagger}d + c_{-1}^{\dagger}d) + \mathrm{h.c.}\right]
\end{equation}
\begin{equation}
H_{ext} =  -J\sum_{i=-L+1}^{L-1}(c_i^{\dagger}c_{i+1}+\mathrm{h.c.}) ~-~ J'\left(c_0^{\dagger}d ~+~ \mathrm{h.c.}\right).
\end{equation}
\begin{equation}
H_{end} = -J\sum_{i=1}^{L-1}  (c_i^{\dagger}c_{i+1}+\mathrm{h.c.})  ~-~ J'\left(d^{\dagger}c_1 ~+~ \mathrm{h.c.}\right) 
\end{equation}
The impurity site is located at $i_{imp}=0$ for embedded and end-coupled RLMs ($d\equiv{c_0}$).  In
the external RLM, it is located at an external site and couples only to site $i=0$ of the chain.
The total system size (bath+impurity) is $\mathcal{L} =2L+1$, $2L$ and $L$ for embedded, external
and end-coupled geometries respectively.  For embedded and external RLMs, one could choose either
periodic or open 1D chains; we work with periodic chains.  For infinite chains or with open boundary
conditions, the embedded Hamiltonian can be mapped onto the endpoint Hamiltonian with a rescaled
$J'$.  

We will restrict to half-filling, which corresponds to zero chemical potential.  The Hamiltonians
above are written for the case where the impurity level is tuned to the chemical potential, 
$\epsilon_d=0$, and hence the $\epsilon_dd^{\dagger}d$ term is omitted.  The effect of a detuning
term will be explored in Section
\ref{sec_detuning}.

\section{\label{sec_greens_function}   Two-point correlator: analytic results}

\begin{figure}
\centering \includegraphics[width=0.8\linewidth]{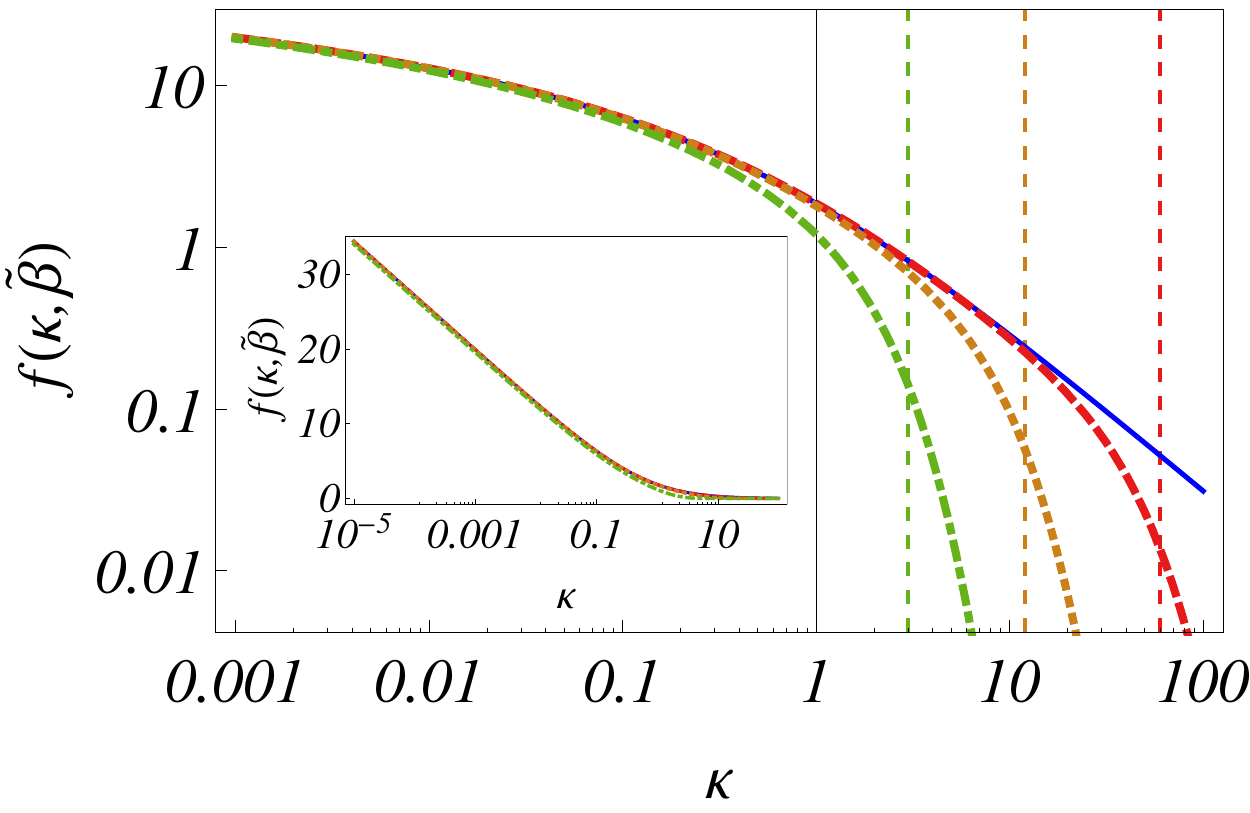}
\caption{\label{fig_analyticT}
Scaled envelope of two-point correlator $f(\kappa, \tilde{\beta})$ (defined in
Eqs.\ \ref{gdc3},\ref{gdc4}) as a function of scaled distance $\kappa$, in log-log scale.  We show
both zero temperature (blue solid line) and finite temperatures: $\tilde{\beta}=60$ (red dashed
line), $\tilde{\beta}=12$ (orange dotted line), $\tilde{\beta}=3$ (green dash-dotted line).  The
$\kappa=1$ vertical line indicates the crossover lengthscale (size of screening cloud).  The
finite-temperature curves deviate sharply from $f(\kappa)$ at distances larger than
$\kappa=\tilde{\beta}$ (shown with dashed lines for the three $\tilde{\beta}$ values).
Inset: $f(\kappa)$ plotted in log-linear scale.  The $-\ln\kappa$ behavior within the cloud
($\kappa<1$) is clear from the constant slope.
}
\end{figure}

In this section we present the essential features of the screening cloud using analytic results for the 
 $\xpct{d^{\dagger}c_x}$ correlator.  This can be derived at finite temperature using standard  means, yielding:
\begin{eqnarray}
\left\langle d^{\dagger}c_{i}\right\rangle =
-J'\frac{1}{L}\sum_{k}\int d\nu\, A_{dd}(\nu)\frac{n_{f}\left(\nu\right)-n_{f}\left(\varepsilon_{k}-\mu\right)}{\nu-\left(\varepsilon_{k}-\mu\right)}e^{-ikr_{i}},\label{gdc1}
\end{eqnarray}
where $A_{dd}(\nu)$ is the spectral function of the impurity and
$n_{f}\left(\nu\right)=1/\left(1+e^{\beta\nu}\right)$ is the Fermi function at temperature
$T=1/\beta$. For completeness we provide explicit details of the derivation of Eq.(\ref{gdc1}) in
the Appendix.

Assuming that the dispersion relation of the bath electrons remains
linear within an energy window $\Lambda$ around the Fermi level, i.e., 
$(\epsilon_{k}-\mu)\simeq v_{F}\left(\left|k\right|-k_{F}\right)$
for $\left|\epsilon_{k}-\mu\right|\lesssim\Lambda$,  and that $T/\Lambda,\Gamma/\Lambda\ll1$
(where $\Gamma$ is the characteristic energy broadening of  the
impurity spectral function), Eq.(\ref{gdc1}) can be approximated
by 
\begin{eqnarray}
\left\langle d^{\dagger}c_{i}\right\rangle =-\frac{J'\rho_0}{\pi} 
Re \left[e^{ik_{F}r_{i}}\int d\varepsilon\int d\nu\,
  A_{dd}(\nu)\frac{n_{f}\left(\nu\right)-n_{f}\left(\varepsilon\right)}{\nu-\varepsilon}e^{i\frac{\varepsilon
      r_{i}}{v_{F}}}\right].   \label{gdc2}
\end{eqnarray}
In line with the above approximations the density of states of the bath electrons in the absence of
the coupling is taken to be constant within the $\Lambda$-window:
$\rho\left(\nu\right)\simeq\rho_0\Theta\left(\Lambda-\left|\nu-\mu\right|\right)$.

For the RLM in the wide-band limit the impurity spectral function can be approximated
by 
\begin{eqnarray} 
A_{dd}(\nu)=\frac{1}{\pi}\frac{\Gamma}{(\nu-\epsilon_{d}+\mu)^{2}+\Gamma^{2}} , 
\label{eq:Add}
\end{eqnarray}
with $\Gamma=\pi J'^{2}\rho_0$ corresponding to the hybridization width (see Appendix). In the
following we assume that the resonance condition $\epsilon_{d}=\mu$ is always fulfilled. In this
case Eq.~(\ref{gdc2}) further simplifies to
\begin{eqnarray}
\left\langle d^{\dagger}c_{i}\right\rangle  & = & \frac{J'\rho_0}{\pi}{\rm
  Re}\left[e^{ik_{F}r_{i}}f\left(\kappa,\tilde{\beta}\right)\right] \label{gdc3}
\end{eqnarray}
where $\kappa=\frac{r_{i}}{\xi}$, $\xi=\frac{v_{F}}{\Gamma}$,
$\tilde{\beta}=\frac{\Gamma}{T}$, and with $f$ given by 
\begin{eqnarray}
f\left(\kappa,\tilde{\beta}\right) & = & \pi\int_{-\infty}^{\infty}dx\frac{x{\rm cos}(\kappa x)+{\rm
    sin}(\kappa x)}{(x^{2}+1)(1+e^{\tilde{\beta}x})} \label{gdc4}
\end{eqnarray}
 At zero temperature, defining $f\left(\kappa\right)=f\left(\kappa,\beta\to\infty\right)$,
one obtains the asymptotic forms 
\begin{eqnarray}
f\left(\kappa\right) & \simeq & -\frac{\pi}{\kappa}; \quad  \kappa\to\infty\\
f\left(\kappa\right) & \simeq & -\pi\left[\ln(\kappa)+\gamma\right]; \quad  \kappa\to0
\end{eqnarray}
with $\gamma$ the Euler constant.
The scaling function $f(\kappa,\tilde{\beta})$ is plotted
in Figure \ref{fig_analyticT}.  For finite $\tilde{\beta}$ one can identify $\kappa_{T}=\tilde{\beta}$
such that $f(\kappa,\tilde{\beta})\simeq f(\kappa)$
for $\kappa\ll\kappa_{T}$. 

We note that the broadening of the spectral function, $\Gamma=\pi\rho_0J'^{2}$, acts as the
characteristic energy scale.  We therefore identify this as the analog of the Kondo temperature for
the RLM, and denote it as $T_{SC}$ to highlight the connection to screening cloud formation The
characteristic length scale is $\xi=v_{F}/T_{SC}$.

\subsection{\label{sec_Anderson_comments} Implications for the Anderson model}

As it stands, Eq.~(\ref{gdc1}) is valid not only for the RLM but also for the Anderson impurity
model.  Thus we have a prediction for the correlators $\xpct{d_{\sigma}^{\dagger}c_{x\sigma}}$ in
the SIAM.  (Here $\sigma$ is $\uparrow$ or $\downarrow$.)  If $\Gamma_{SIAM}$ is the broadening of
the spectral function in the SIAM, the behavior of this correlator will be $\sim-\ln(x/\xi_{SIAM})$
for $x<\xi_{SIAM}$ and $\sim{x}^{-1}$ for $x>\xi_{SIAM}$, where $\xi_{SIAM}=v_F/\Gamma_{SIAM}$.  One
expects these functional forms to hold at low temperatures, even if the spectral function is not an
exact Lorentzian and even if the spectral function has temperature dependence.

The correlator usually used for describing the screening cloud for the Anderson model is not
$\xpct{d_{\sigma}^{\dagger}c_{x\sigma}}$ but the spin-spin correlator \cite{Goth_Assaad,
  Holzner-etal_PRB09}.  For $U=0$, Wick's theorem implies this to be proportional to the square of
$\xpct{d_{\sigma}^{\dagger}c_{x\sigma}}$.  Thus we can expect
$\xpct{\vec{S}_{\mathrm{imp}}\cdot\vec{s}_x}$ to behave like $[\ln{x}-\ln\xi_{SIAM}]^2$ within the
screening cloud for small values of $U$.  We have found that the data in Ref.~\cite{Goth_Assaad} is
qualitatively consistent with this prediction for interactions as large as $U\sim2$.

\section{\label{sec_finite_size} Two-point correlators on finite-size lattice RLM's}

\begin{figure}
\centering \includegraphics[width=0.7\linewidth]{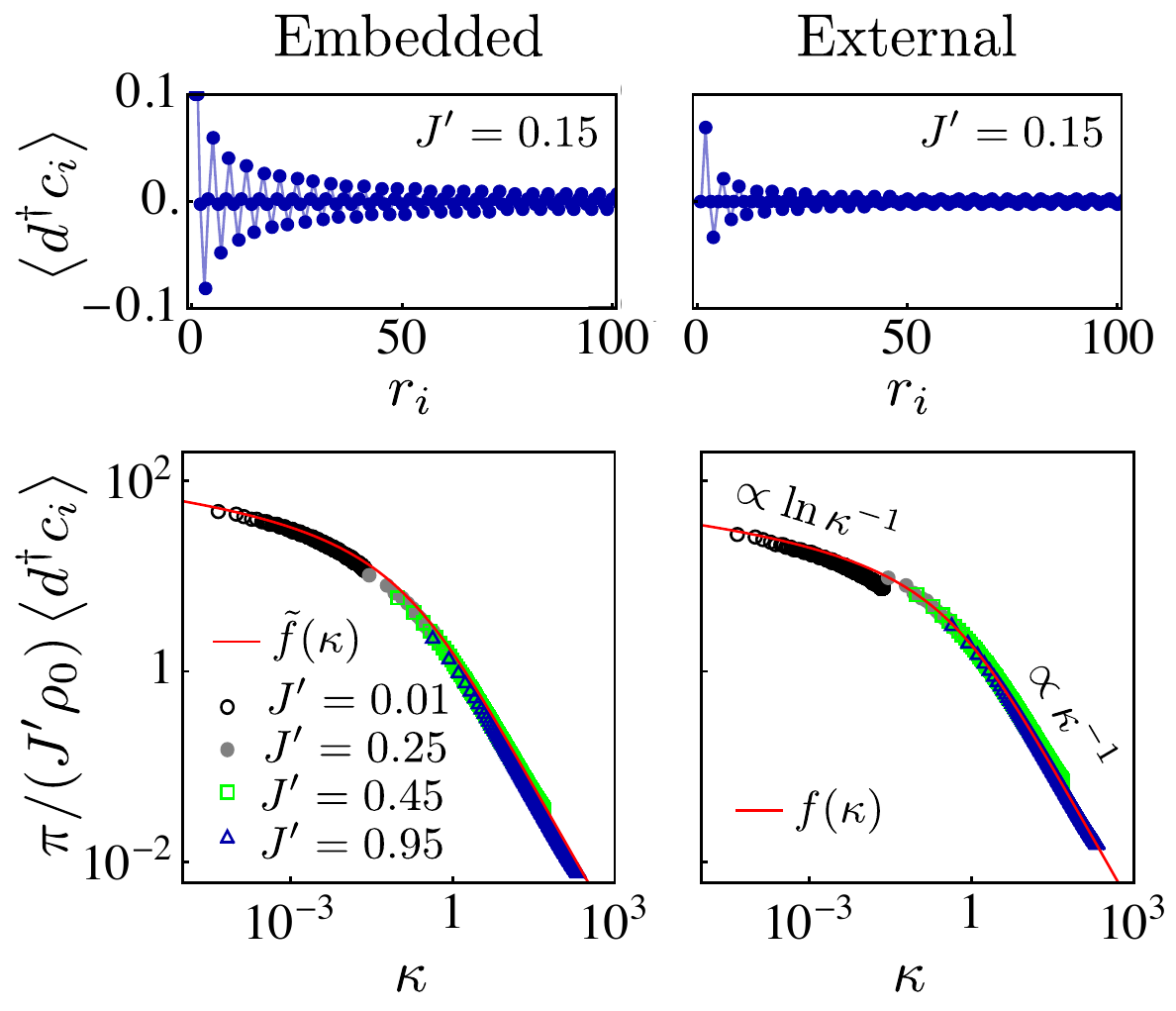}   
\caption{ \label{fig_correlation_function} 
Upper panels: Correlator $\xpct{d^{\dagger}c_i}$ plotted against distance $r_i$ from the impurity
site for finite-size RLM systems in the embedded and external geometries. 
System sizes and number of particles are $\mathcal{L}=2003$, $N=1001$ for the embedded and $\mathcal{L}=2000$, $N=1000$ for the external geometries. 
Lower panels: Scaled envelope of $\xpct{d^{\dagger}c_i}$ against scaled distance $\kappa$ plotted for different values of $J'$. 
The external case (right) is compared with $f(\kappa)$ function.The embedded case (left) is compared with $\tilde f(\kappa) \simeq 2.0
 f( 3.2  \kappa)$. 
}
\end{figure}

In this section we present numerical results for the equal-time correlator $\xpct{d^{\dagger}c_i}$
characterizing the spatial structure of the screening cloud in RLMs on finite chains of length
$\mathcal{L}$ for the three geometries introduced in Section \ref{sec_geometries} at zero and finite
temperature.  The analytic predictions of Section \ref{sec_greens_function} are directly applicable to
the external geometry, but we will show that the predicted scaling matches the end-coupled and
embedded cases with the use of simple scaling factors.  Boundary effects are found to be different
for the three geometries.

For our numerical calculations we consider systems of size $\simeq10^3$ at half-filling with
even (odd) number of total lattice sites for end-coupled and external (embedded) RLMs.

\subsection{Zero temperature} 

The single-particle correlation functions $\xpct{d^{\dagger}c_i}$, at temperature $T=0$, are shown
in Figure \ref{fig_correlation_function} for different geometries.  The correlators
oscillate as $\sim {\rm cos}(k_F r_i)$ with distance $r_i$ from the impurity site.  Since we are at
half-filling, the Fermi momentum $k_F$ is commensurate with the lattice spacing, so the envelope of
oscillations can be obtained by plotting $|d^{\dagger}c_{( i= 2n )}|$, where $n$ is an integer.  
The lower row of Figure \ref{fig_correlation_function}
shows the envelopes obtained in this way.

For these sizes ($\simeq 10^3 $),  the envelope for a single value of $J'$ follows only a small
part of the scaling curve $f(\kappa)$.  
These individual curves show finite-size deviations near the system boundaries.  
The curves for many $J'$ together reconstruct very well
the full scaling curve $f(\kappa)$ for the external geometry.  
For the embedded and end-point geometries the scaling collapse still occurs, 
but the scaling function and its argument have to be rescaled, $\tilde{f}(\kappa) = Af(c\kappa)$.

The collapse of the envelopes for different $J'$ onto the single curve confirms the
existence of the finite screening length scale $\xi$ in all three realizations of RLMs, conjectured
from analytical calculation for the external RLM.

For small $J'$ ($= 0.01J$), the screening length $\xi$ is much larger
that the system size $\mathcal{L}$ and the $ r_i^{-1}$ behavior of the free-fermionic correlator is
absent as the impurity is not completely screened within the length of the system.  On the other
hand, for large $J'$ ($= 0.95J$), $\xi \ll \mathcal{L}$ and the correlation function behaves mostly as
$r_i^{-1}$ as the impurity gets screened over a very small distance.

\subsection{\label{sec_log_vs_powerlaw} Logarithm versus power law within the screening cloud}

\begin{figure}
\centering \includegraphics[width=0.8\linewidth]{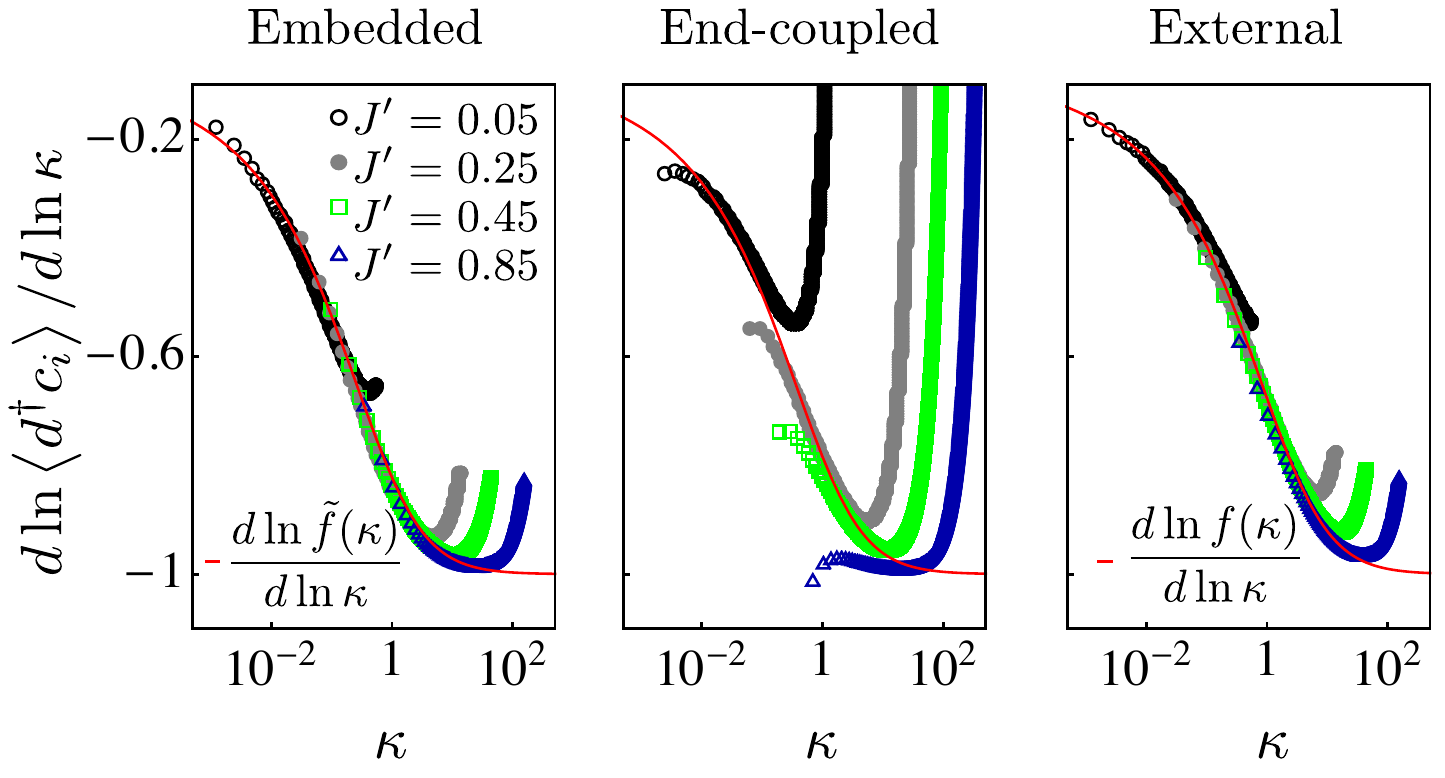}
\caption{\label{dcderiv}
Plot of logarithmic derivative  $d \ln\xpct{d^{\dagger}c_i}/d{\ln \kappa}$ 
as a function of $\kappa$ for embedded,
end-coupled and external RLMs. For the external case the solid (red) line corresponds to the derivative of $\ln f(\kappa)$ obtained from analytical 
calculation. For the other two cases the envelope is seen to behave as $ A f( c \kappa )$ with $A$ and $c$  constants of order $1$. The fitted values of $c$ are $c\simeq3.2$ (embedded) and $c\simeq2.3$ (end-coupled). 
The system sizes $\mathcal{L}$ are same as in Fig.\ \ref{fig_correlation_function}}.
\end{figure}

In the isotropic Kondo model, impurity-bath spin-spin correlators have been reported to show a
crossover, at the screening length-scale, between power-law behaviors of different exponents
\cite{Borda}. In contrast, in the RLM, the correlator does not become a power-law of smaller exponent
within the screening cloud, but rather becomes a logarithm.  We highlight this in Figure
\ref{dcderiv} by plotting the logarithmic derivative $\partial \ln\xpct{d^{\dagger}c_i}/\partial
(\ln \kappa)$ as a function of $\kappa$ from the numerical data for all three geometries.  We also
compare with the analytical predictions.

When the dependence is a power-law, the logarithmic derivative is the corresponding exponent.
Therefore it converges to $-1$ at large distances.  At small distances within the screening cloud
($\kappa<1$), the logarithmic derivative increases continuously toward zero as $\kappa$ decreases.

The numerical curves for each $J'$ shows finite size effects as the system boundary is approached, 
but otherwise the curves collapse onto the analytical prediction in the external case.  In the
other two geometries the rescaled form $\tilde{f}(\kappa) = Af(c\kappa)$ has to be used.

\subsection{Finite temperature}

\begin{figure}
\centering \includegraphics[width=0.8\linewidth]{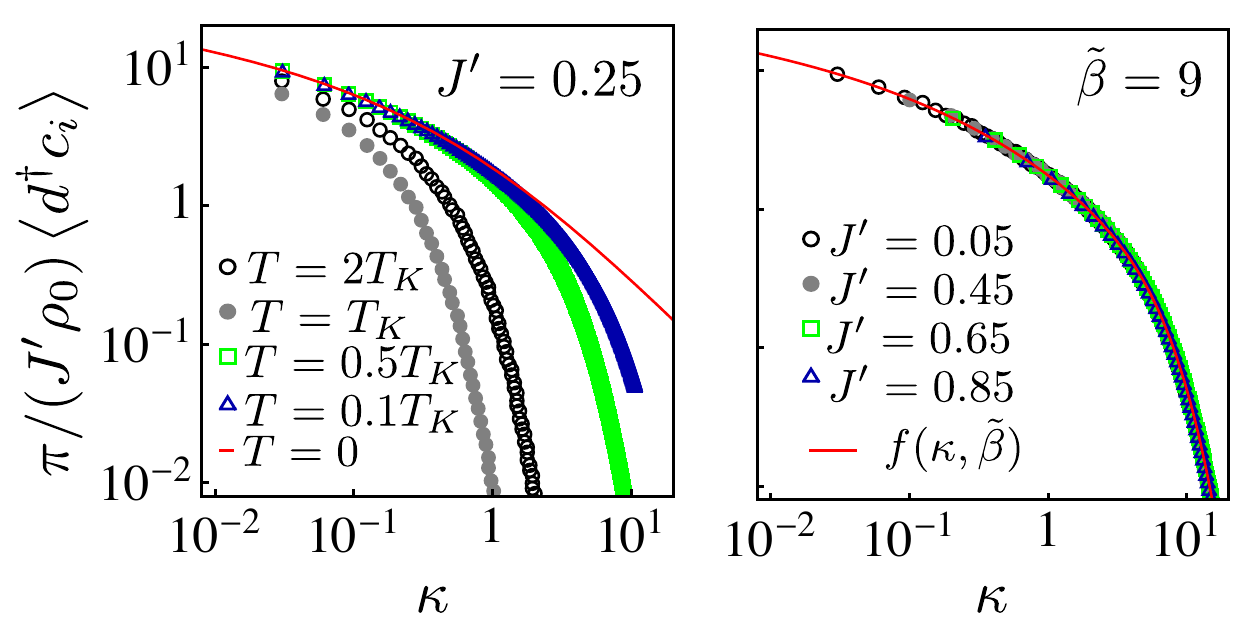}
\caption{ \label{fig_correlator_1}  
Envelope of the correlation function $\xpct{d^{\dagger}c_i}$ plotted against the scaled distance $\kappa$.
Left panel: Fixed $J'=0.25J$ and different values of the temperature. The solid (red) line corresponds to $f(\kappa)$. 
Right panel: Fixed rescaled temperature $\tilde{\beta}=9$ and different values of $J'$. The solid (red) line corresponds to $f(\kappa,\tilde{\beta}=9)$.
}
\end{figure}

Next, we consider the correlation functions at finite tempertaure.  The screening length $\xi$
corresponds to the temperature scale $T_{SC}=v_F/\xi$, above which the screening of impurity by
conduction electrons is thermally destroyed.  In Fig.\ref{fig_correlator_1}-(left), the envelope
of oscillations are shown for the external geometry, for several different temperatures.  
Finite temperature induces another length scale $\xi_T=v_F/T$, the thermal length scale.  
The  behavior of $\xpct{d^{\dagger}c_i}$ is not much affected by the temperature
in regions $r_i<\xi_T$ while it shows an exponential decay $\sim e^{-r_i/\xi_T}$ for $r_i>\xi_T$.  
Plots of the rescaled envelope as a function of $\kappa$, keeping $\tilde{\beta}=T_{SC}/T$ fixed, shown
in Fig.\ref{fig_correlator_1}-(right), confirm the predicted scaling form
$f(\kappa,\tilde{\beta})$ of the correlator.  In Fig.\ref{fig_correlator_1} we present finite temperature
numerical results only for the external RLM. Similar features are observed in the other two
geometries.

\section{\label{sec_detuning} RLM with on-site potential}

\begin{figure}
\centering \includegraphics[width=0.9\linewidth]{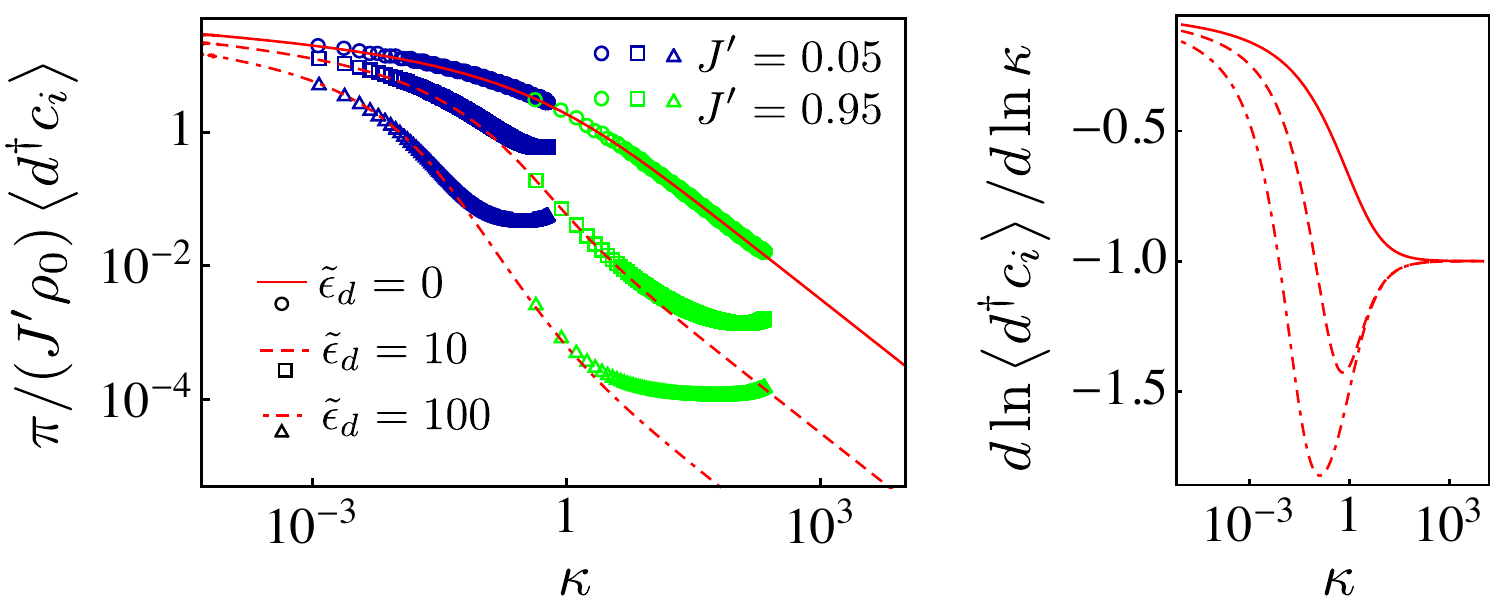}
\caption{\label{finiteV} Effect of mismatch  $\epsilon_d$ between impurity site energy and bath chemical
  potential .  Left: Scaled envelope of $\xpct{d^{\dagger}c_i}$ as a function of $\kappa$ for
  $\tilde{\epsilon_d}=0,10,100$.  Full curves show analytic function $f(\kappa,
  \tilde{\epsilon_d})$, while the dots are numeric data for $\mathcal{L}=2000$.  The
  full scaling curve can be constructed with a combination of several $J'$ values; here we show only
  two values.  Right: Plot of logarithmic
  derivative of $f(\kappa, \tilde{\epsilon_d})$ with same values of $\tilde{\epsilon_d}$ as in left
  panel.}
\end{figure}

So far we have considered the on-site energy of the impurity level $\epsilon_d$ to be the same as the chemical
potential $\mu=0$ of the fermionic bath.  In this section we consider the effect of finite
$\epsilon_d$ on the screening cloud.  
We focus on the external RLM at half-filling and zero temperature; the relevant Hamiltonian is
$H_{ext}+\epsilon_dd^{\dagger}d$.  

The analytical expression for $\xpct{d^{\dagger}c_i}$ in the presence of impurity detuning can be
obtained using the Green's function method described in Section \ref{sec_greens_function}:
\begin{equation}
\xpct{d^{\dagger}c_i}= \frac{J'\rho_0}{\pi}{\rm
  Re}\left[e^{ik_{F}r_{i}}f\left(\kappa,\tilde{\epsilon_d}\right)\right] 
\end{equation}
where 
\begin{equation}
f\left(\kappa, \tilde{\epsilon_d} \right)= \pi \int_{-\infty}^0 dx \frac{e^{i\kappa
    x}}{x-\tilde{\epsilon_d}+i}
\end{equation}
Here $\tilde{\epsilon_d}=\epsilon_d/\Gamma$ is the scaled impurity energy. A plot of this analytical
function together with numerical results are shown in Figure \ref{finiteV}-(left).  The $\kappa^{-1}$
behavior of the correlation function is still present for regions $\kappa>1$ outside the
cloud. However, another region, with non-logarithmic behavior, develops
within the cloud ($\kappa<1$). This region expands from the exterior of the cloud towards its
center at the impurity site and increases with increasing $\epsilon_d$, thus destroying the
characteristic logarithmic behavior of the screening cloud.  The loss of the logarithmic region is
highlighted in Figure \ref{finiteV}-(right) using the logarithmic derivative.

\section{\label{sec_entanglement}  Entanglement entropy}

\begin{figure}
\centering \includegraphics[width=0.9\linewidth]{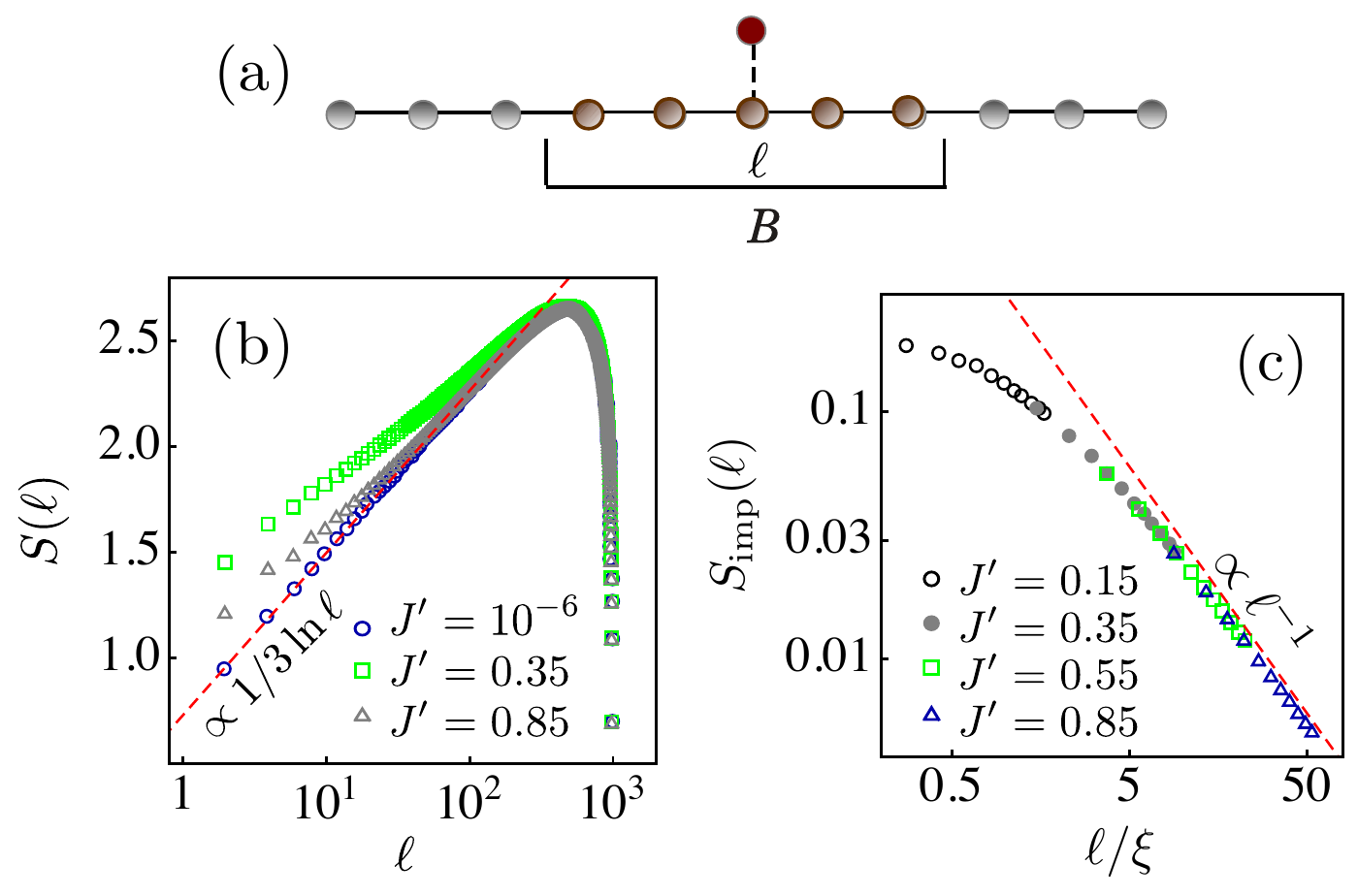}
\caption{ \label{extentropy1} (a) Schematic diagram of subsystem B of length $\ell$, including
  the impurity site in external RLM.  
(b) Entanglement entropy $S(\ell)$ between subsytem B and the rest as a function of
 subsystem size  $\ell$  for $\mathcal{L}=1000$ and several $J'$ values. 
 For very small $J'=10^{-6}J$, one observes the expected logarithmic behavior $ S(\ell) \simeq \frac{1}{3}\ln{\ell}+$const.  for $\ell\ll\mathcal{L} $. 
(c) Impurity entanglement entropy $S_{imp} = S- S _{J'=0}$ 
as a function of  $\ell/\xi$
 for several values of $J'$, keeping the value of $\ell/\mathcal{L}$ fixed to 1/4. A scaling collapse is observed for different values of $J'$.
 For large $\ell$, $S_{imp}$ vanishes as $\ell^{-1}$. }
\end{figure}

Having characterized the screening cloud using impurity-bath correlators in most of this work, in
this section we focus on a different quantity.
We consider the entanglement entropy of a subsystem ($B$) of length $\ell$, including the impurity site
at its center, with the rest of the system ($A$) .  The entanglement entropy is defined as $S=-{\rm
  Tr}_{B}[\rho_B{\rm ln}\rho_B]$, where $\rho_B ={\rm Tr}_{A}\rho$ is the reduced density
matrix of subsystem $B$, obtained by tracing over the $A$ degrees of freedom.  For free
fermionic systems like the RLM, the entanglement entropy can also be expressed as $S=-\sum_i [\nu_i{\rm
    ln} \nu_i+(1-\nu_i) {\rm ln} (1-\nu_i)]$, where $\nu_i$'s are the eigenvalues of one particle
correlator $[C_{ij}]=[\xpct{c_i^{\dagger}c_j}]$, $i,j \in~B$. 

As shown in Ref.\ \cite{Laflorencie} for spin chains, such block entanglement entropies exhibit
signatures of the screening length scale. The impurity entanglement entropy is defined as
\begin{equation} 
S_{\rm imp}(\ell)=S(\ell)-S_{J'=0}(\ell).
\end{equation}
 This quantity is expected to follow a scaling form, i.e. to depend only on  
the ratio $\ell/\xi$, provided that  $\ell / \mathcal{L}$ is constant or $\ell \ll \mathcal{L}$.

We present here the numerical results for entanglement entropy of external RLM; the entanglement
entropy for endpoint/embedded geometries has very similar features and has been presented recently
in Ref.\ \cite{SaleurVasseur_PRB13}.
Fig.\ \ref{extentropy1}-(b) shows $S(\ell)$ for several values of $J'$.  For $J'=0$, which
corresponds to the system without impurity, the entanglement entropy has the form
$S_{J'=0}(\ell)=\f{1}{3}\ln{\ell}+ const.$ for $\ell \ll \mathcal{L}$, confirming the
prediction from conformal field theory with central charge $c=1$.  Fig. \ref{extentropy1}-(c) shows
$S_{\rm imp}(\ell)$ as a function of the scaled variable $\ell/\xi$ for different values of $J'$,
keeping $\ell/\mathcal{L}$ fixed at $1/4$.  The data for different $J'$ collapses onto a single
curve.

\section{\label{sec_next_nearest_neighbor}  "Non-Fermi liquid" behaviour from an unconventional bath}

\begin{figure}
\centering \includegraphics[width=0.9\linewidth]{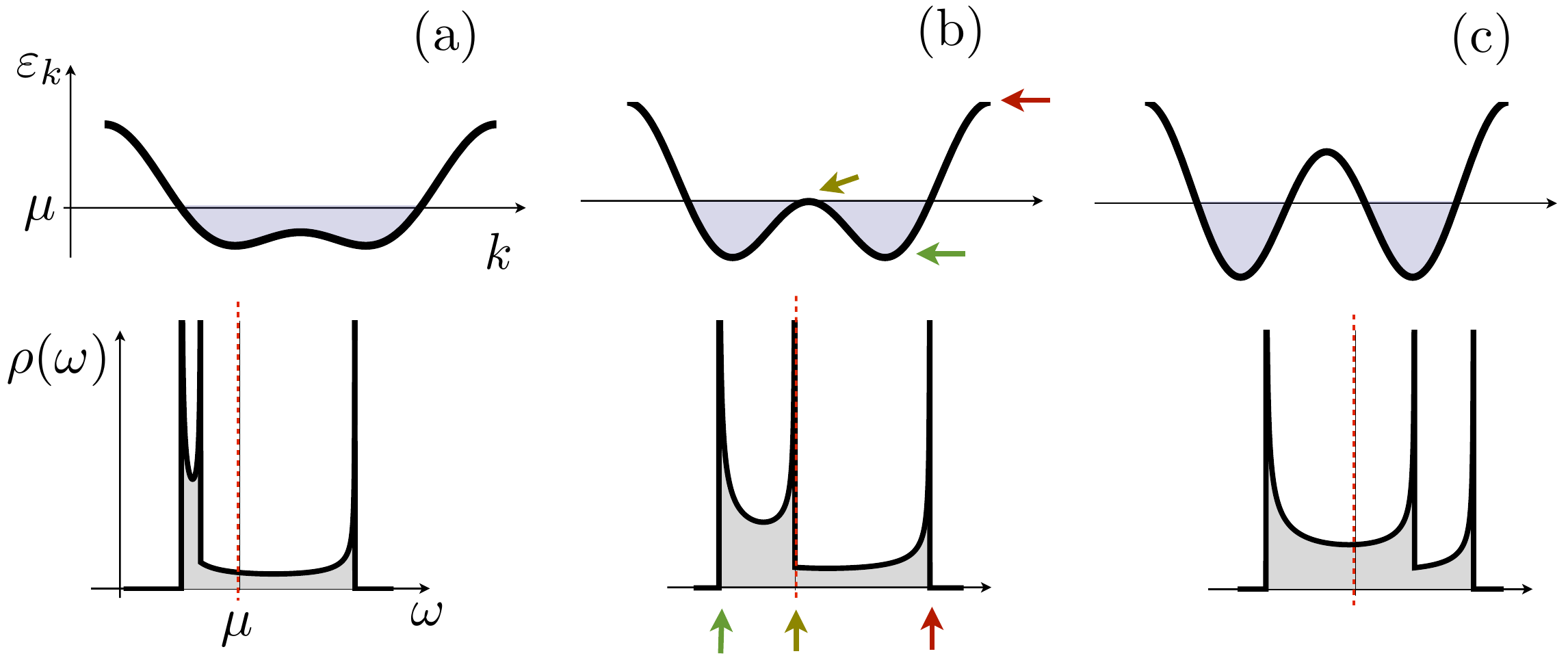}
\caption{ \label{FigNNN} Dispersion relations and the respective densities of states, for three
  cases with different Fermi surface geometry.  The cases shown correspond to the Hamiltonian
  $H_{NN}$ of Eq.\ (\ref{Hnnn}) with $J'=0$ and: (a) $J_{NN} = -0.5 J$; (b) $J_{NN} = - J$; (c)
  $J_{NN} = -3 J$. The finite density of states at the Fermi level of cases (a) and (c) induces the
  scaling of $\xpct{d^{\dagger}c_i}$ given in Eqs.\ (\ref{gdc3},\ref{gdc4}). For case (b) where
  $\rho(\omega \to \mu ) \propto \left|\omega\right|^{-1/2} \Theta(-\omega)$ the impurity cloud
  follows Eqs.\ (\ref{dc2},\ref{dc3}).  }
\end{figure}

The treatment of section \ref{sec_greens_function} is for a single set of Fermi points such as the
case shown schematically in Fig.\ref{FigNNN}-(a).  This analysis can be generalised to the case of
multiple sets of Fermi points such as the case shown in Fig.\ref{FigNNN}-(c).  In this case the
envelope of the impurity cloud still follows the scaling form of Eq.(\ref{gdc4}) with the density of
states getting a contribution from each one of the Fermi point pairs: $\rho_0 \simeq \pi^{-1}(\left|
v_{F,1} \right|^{-1} + \left| v_{F,2} \right|^{-1}) $, where $v_{F,i} $ is the velocity at Fermi
point $i$. We assume a parity symmetric dispersion relation $\varepsilon_k = \varepsilon_{-k}$.

A qualitatively different behaviour is obtained when the density of states at the Fermi level
diverges, a situation that may occur when the 1D Fermi "surface" changes its
topology. Fig.~\ref{FigNNN}(b) shows this situation arising at the transition between cases (a) and
(c). In this section we extend our analysis to such a case. We show below that with such an
unconventional bath the asymptotic behavior of the correlator $\xpct{d^{\dagger}c_i}$ is
qualitatively different from the regular Fermi liquid expectation $\kappa^{-1}$. The nature of this
bath also introduces a new length/energy scale.

In a situation like Fig.~\ref{FigNNN}(b), there is a momentum value, assumed without loss of
generality to be $k=0$, where the dispersion relation is quadratic: $|\epsilon_k-\mu| \simeq a_F
k^2$.  It follows that the density of states of the fermionic bath diverges as $\rho(\omega \to \mu
) \simeq 1/(4\pi) \left| a_F (\omega-\mu) \right|^{-1/2} \Theta(\mu-\omega) + \rho_0 $ (see
Fig.\ref{FigNNN} (b) lower panel) where $\rho_0=1/(\pi |v_F|)$ is the contribution from the regular
Fermi points.  The derivation of the correlator $\xpct{d^{\dagger}c_i}$ follows closely the one of
case (a) (further details are provided in the Appendix) and yields
\begin{eqnarray}
\xpct{d^{\dagger}c_i} = \frac{J' \rho_0}{\pi} g(\kappa,s) 
 \label{dc2}
\end{eqnarray}
with $s=v_F/\sqrt{\Gamma a_F}$. 
The scaling function $g$ is given by
\begin{eqnarray}
 g(\kappa,s) &=& {\rm Re}  \left[ e^{ik_F r_i}  \int_{0}^{\infty}  
dy   \left\{
\frac{ e^{-i \kappa  y} \Gamma (0,-i y \kappa ) \left(1+\frac{s}{2\sqrt{y}}\right)}{
  y^2+(1+\frac{s}{2\sqrt{y}})^2    }  \right. 
 + \frac{  e^{i \kappa  y} \Gamma (0,i y \kappa )  }{ (y-\frac{s}{2\sqrt{y}})^2+1}\right\}         \nonumber \\
&& \left.  
 +\pi s \int_0^{\infty} du \frac{e^{-u s \kappa}}{(u^2-\frac{s}{2 u})^2+1} \, \right] 
 \label{dc3} 
\end{eqnarray}
with $\Gamma (a,z)$ the incomplete gamma function.
Note that $g(\kappa, s)$ reduces to $f(\kappa)$ in Eq.(\ref{gdc4}) for $s = 0$.
The dimensionless parameter $s$ quantifies the effect of the diverging density of states. 

For finite $s$ the function $g(\kappa, s)$ is composed of two different contributions arising from
the regular Fermi points (first term inside the square brackets in Eq.\ref{dc3}) and from the
special point at $k=0$ (second term inside the square brackets).

\begin{figure}
\centering \includegraphics[width=0.8\linewidth]{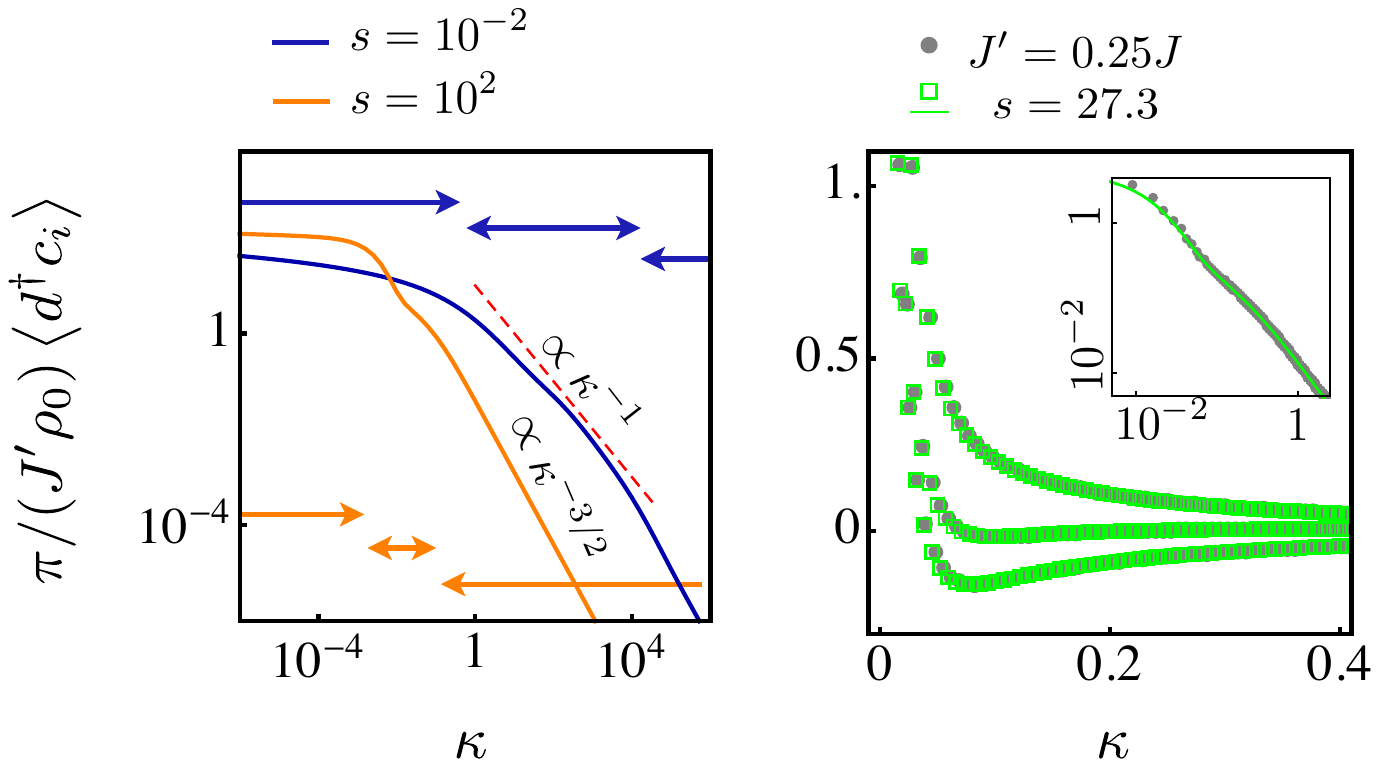}
\caption{ \label{FigNNN2} 
Screening cloud for bath with diverging density of states,  case (b) of Fig.~\ref{FigNNN}.  
Left: Envelope of $\xpct{d^{\dagger}c_i}$ plotted against the scaled distance $\kappa$, using
Eq.~(\ref{dc3}) .  A small $s$
(blue/dark-solid) and a large $s$ (orange/light-solid) case is shown.  The different scaling regions
are indicated by arrows near the top (for $s=10^{-2}$) and bottom ($s=10^2$) of the panel.
Right: correlator $\xpct{d^{\dagger}c_i}$ for the $H_{NN}$ Hamiltonian with
$J=-J_{{NN}}=1$ and $J'=0.25J$, compared with Eq.(\ref{dc3}) with corresponding value
$s\approx27.3$.  In the main panel, the full oscillatory behaviour is shown and not just the
envelope.
The inset shows a comparison of the envelope.  
}
\end{figure}

For very small values of $s\ne 0$ the contribution from the second term in Eq.(\ref{dc3}) is
negligible.  The envelope of the correlator $g(\kappa, s)$ exhibits three different scaling regions
(see Fig.~\ref{FigNNN2} - left panel) arising only from the first term: 
\\ (i) $g(\kappa, s) \propto -\ln\kappa$ for $\kappa\ll 1$,
\\ (ii) $g(\kappa, s) \propto \kappa^{-1}$ for $1\ll\kappa\ll s^{-2}$,  and
\\ (iii) $g(\kappa, s) \propto \kappa^{-3/2}$ for $\kappa\gg s^{-2}$.  
\\
The intermediate
$\sim\kappa^{-1}$ region is barely visible for $s=10^{-2}$ (Fig.~\ref{FigNNN2}) but becomes clearer
and more extended at even smaller $s$.  

For large $s$, the contribution from the special ($k=0$) point is large and shows a crossover from a
logarithmic $g(\kappa, s) \propto \ln s\kappa$ to power-law dependence $g(\kappa, s) \propto
(s\kappa)^{-3}$.  The first term in Eq.(\ref{dc3}) also shows two different
scaling behaviors: $g(\kappa, s) \propto \ln \kappa$ for small $\kappa$ and $g(\kappa, s) \propto
\kappa^{-3/2}$ for large $\kappa$. As a result the scaling function $g(\kappa, s)$, composed of these
two contributions, does not show the $\kappa^{-3}$ behavior as the logarithmic dependence coming
from the first term overcomes this faster decaying power-law. One then finds three different scaling
regions for large values of $s$ (Fig.~\ref{FigNNN2} left panel):
\\ (i) $g(\kappa, s) \propto \left(-\ln\kappa + c_1\ln{s} \right)$ for $\kappa\ll s^{-4/3}$, 
\\ (ii) $g(\kappa, s) \propto -\ln\kappa$ for $s^{-4/3} \ll\kappa\ll s^{-2/3}$, and 
\\ (iii) $g(\kappa, s) \propto \kappa^{-3/2}$ for $\kappa\gg s^{-2/3}$.  
\\ The first and second regions both have dominant $-\ln\kappa$ behaviors, but they are shifted by a
constant $\sim\ln{s}$.  This shift is visible (Fig.~\ref{FigNNN2} left panel) as a jump between the
two regions, around $\kappa\sim s^{-4/3}$.

In order to illustrate  our analytical findings we consider a minimal model with next nearest neighbor hopping of the 
lattice electrons:
\begin{equation}
 H_{NN} =H_{ext} -  J_{NN} \sum (c_i^{\dagger}c_{i+2}+ h.c.) \label{Hnnn}
\end{equation}
where the filling fraction is determined by fixing the chemical potential at $\mu=0$ and the dispersion relation at $J'=0$ is given by $\varepsilon_k= - 2 J \cos{k} - 2 J_{NN} \cos{2 k} $.
Depending on next nearest neighbor hopping parameter $J_{NN}$ the energy spectrum
of the bath can be categorized in three different cases as shown in Fig.\ref{FigNNN} : 
(a) for $-J_{NN} <J$, 
(b) for $-J_{NN} =J$ ,
(c) for $-J_{NN} >J$ .

Fig.\ref{FigNNN2} shows $\xpct{d^{\dagger}c_i}$ computed for the (b) case with $J=-J_{NN}=1$
and $J'= 0.25J$.  The regular Fermi momentum situated at $k_F=2\pi/3$ sets an oscillatory behaviour
in real space with period 3. This is visible in Fig.\ref{FigNNN2}-right panel as three
branches. This set of parameters corresponds to $s\approx27.3$.  The numerical data is in excellent
agreement with $\xpct{d^{\dagger}c_i}$ calculated from Eq.(\ref{dc3}) with $s=27.3$
(Fig.\ref{FigNNN2}-right, open green points).

The envelope of the correlator can be obtained considering $\xpct{d^{\dagger}c_i}$ for $i=3n$ (with
$n$ integer) in the same spirit as in section \ref{sec_greens_function}. The numerically obtained
envelope and the analytic prediction are shown in the inset of Fig.\ref{FigNNN2}-right to follow the
predicted $\kappa^{-3/2}$ asymptotic behavior.

\section{\label{sec_d_larger_than_one} Higher dimensional baths}

For a resonant level embedded in an electronic bath of dimensionality $D>1$, the angular dependence
of the Fermi surface has to be taken into account.  In general, a higher-dimensional Fermi surface
will be anisotropic, e.g., the $D=2$ square lattice at half filling has a square-shaped
Fermi surface.  Below we provide some expressions for the $\langle{d^{\dagger}c_r}\rangle$
correlator for $D>1$ assuming a circularly/spherically symmetric Fermi surface.

For an isotropic Fermi surface, we linearize the dispersion around the Fermi surface, $\mathbf{k} =
\mathbf{\Omega} (k_F + \varepsilon/v_F)$, with $\mathbf{\Omega}$ a $D$-dimensional unit vector.
Eq.(\ref{gdc1}) then yields
\begin{eqnarray}
\left\langle dc_{r}^{\dagger} \right\rangle &\simeq&
-J'\frac{k_{F}^{D-1}}{\left(2\pi\right)^{D}v_{F}}\int_{-\Lambda}^{\Lambda}d\varepsilon\int d^{D-1}\Omega\int d\nu\times \nonumber \\ 
& &
A_{dd}\left(\nu\right)\exp\left[-i\left(k_{F}+\frac{\varepsilon}{v_{F}}\right)\mathbf{\Omega}\cdot\mathbf{r}\right]
\frac{n_{f}(\nu)-n_{f}(\varepsilon)}{\nu-\varepsilon}
\end{eqnarray}
Here $\Lambda$ denotes a high-energy cutoff of the order of the bandwidth.

The explicit form of resonant level spectral function, $A_{dd}\left(\nu\right)$ given in
Eq.(\ref{eq:Add}), remains unchanged except for the modification of the density of states at the
Fermi level $\rho_0=\frac{k_{F}^{D-1}}{v_{F}\left(2\pi\right)^{D}}\mathcal{A}_{D-1}$ where
$\mathcal{A}_{D-1}=\frac{2\pi^{\frac{D}{2}}}{\Gamma\left(\frac{D}{2}\right)}$ is the area of the $S_{D-1}$
sphere. 
Performing the angular integration and simplifying the previous expression using the explicit form
of $A_{dd}\left(\nu\right)$ one obtains 
\begin{eqnarray}
\left\langle dc_{r}^{\dagger} \right\rangle &\simeq&
-J'\rho_{0}\int_{-\tilde\Lambda}^{\tilde\Lambda} du\,
\frac{h_D\left[\left(\tilde{\epsilon}_{F}+u\right)\kappa\right]+\left(u-\tilde{\epsilon}_{d}\right)g_D\left[\left(\tilde{\epsilon}_{F}+u\right)\kappa\right]}{\left[\left(u-\tilde{\epsilon}_{d}\right)^{2}+1\right]\left(1+e^{\tilde{\beta}u}\right)} \label{eq:dc_23}
\end{eqnarray}
where $\tilde\Lambda=\Lambda/\Gamma$ and  $\tilde{\epsilon}_{F}={\epsilon}_{F}/\Gamma$.  
The functions $g_D$ and $h_D$ are defined as 
\begin{eqnarray}
g_{D}(y) &= (\mathcal{A}_{D-1})^{-1} \int d^{D-1}\Omega\ \exp\left[-i\mathbf{\Omega}\cdot \mathbf{e}_1 y\right] 
\label{eq_def_gD}
\\ h_{D}(y) &= (\mathcal{A}_{D-1})^{-1} \int d^{D-1}\Omega\ \exp\left[-i\mathbf{\Omega}\cdot
  \mathbf{e}_1 y\right]  \, \mathrm{sgn}\left(\mathbf{\Omega}\cdot\mathbf{e}_1 \right)
\label{eq_def_hD}
\end{eqnarray}
with $\mathbf{e}_1$ a fixed unit vector.  Specifically, for $D=2$ one obtains
$g_{2}\left(y\right)=J_{0}(y)$ and $h_{2}\left(y\right)=H_{0}(y)$, respectively the Bessel and
Struve functions of order zero.
For $D=3$, $g_{3}\left(y\right) = \frac{\sin(y)}{y}$ and $h_{3}\left(y\right) =
\frac{\left[1-\cos(y)\right]}{y}$.
For $D=1$, the integrals (\ref{eq_def_gD},\ref{eq_def_hD}) become sums over the Fermi points,
yielding $g_{1}(y)=\cos(y), h_{1}(y)=\sin(y)$; using
these expressions one recovers the on-dimensional solution treated in the previous sections.

At zero temperature and for $\tilde{\epsilon}_{d}=0$,  Eq.(\ref{eq:dc_23}) simplifies to
$\langle{d}^{\dagger}c_{r}\rangle=\frac{J'\rho_{0}}{\pi}T\left(\kappa\right)$ with  
\begin{eqnarray}
T\left(\kappa\right)&=&-\pi\int_{-\tilde\Lambda}^{0}du\,\frac{h_{d}\left[\left(\tilde{\epsilon}_{F}+u\right)\kappa\right]+u g_{d}\left[\left(\tilde{\epsilon}_{F}+u\right)\kappa\right]}{\left[u^{2}+1\right]}. \label{eq:T}
\end{eqnarray}
In the $d=1$ case the oscillatory part of Eq.(\ref{eq:T}) factors out, and the cutoff $\Lambda$ can
be sent to infinity in the remaining expression.  In the $D>1$ cases, unfortunately, this
simplication does not occur except for the limit $\kappa\to\infty$.  Since the cutoff $\Lambda$ is
of the same order as the Fermi energy $\epsilon_F$ which remains within the integral, it is not
consistent to use $\Lambda\to\infty$.  Treating the integral numerically, we have verified that the
large-distance behavior (outside the screening cloud, $\kappa\gg1$) is $T(\kappa)\sim\kappa^{-1}$
for $D=2$ and $D=3$, as it is for the $D=1$ case.  The long-distance behavior reflects an intrinsic
property of the Fermi gas, so this is expected. However, the short-distance behavior (within the
screening cloud, $\kappa\ll1$) is cutoff-dependent and it is difficult to infer a general behavior.
In addition, it should be borne in mind that angular dependence of the Fermi surface, which we have
ignored, may also contribute to making the behavior of $\langle{d}^{\dagger}c_{r}\rangle$
non-universal.

\section{Discussion; Open issues} 

To summarize, we have characterized the spatial structure of the impurity screening cloud for the
resonant level model, using the impurity-bath two-point correlator $\xpct{d^{\dagger}c_x}$ and the
entanglement entropy $S_l$ of a region surrounding the impurity with the rest of the system.
Focusing on 1D baths, we have found that the behavior of the correlator $\xpct{d^{\dagger}c_x}$ is
found to be logarithmic ($\sim-\ln{x}$) within the cloud and power-law ($\sim{x}^{-1}$) outside the
cloud.  The analytic expression in the continuum limit wide-band approximation is provided in
integral form for arbitrary temperatures [Eq.\ (\ref{gdc4})].  The crossover occurs at a length
scale $\xi$ which varies as an inverse square with the coupling $J'$.  The $\xpct{d^{\dagger}c_x}$
envelopes calculated from finite-size lattices with different values of $J'$ reproduce well the full
analytical prediction through a scaling collapse of data.
We have also shown the effect of impurity detuning $\epsilon_d$ from the Fermi energy.  The
screening cloud is robust for small detunings, but at larger detuning it gets destroyed by a new
intermediate-distance regime that grows in spatial extent with the detuning.

The long-distance $\sim{x}^{-1}$ behavior for any $D$ can be understood as the behavior of the
two-point correlator of the undisturbed Fermi gas.  Since the impurity is screened at large
distances, this correspondence is expected.  In the external impurity case, this behavior is
obtained if $\xpct{d^{\dagger}c_x}$ is calculated perturbatively in $J'$.  In contrast, the
short-distance $\sim-\ln{x}$ behavior cannot be explained in the same way.  We are not aware of
alternate arguments or derivations of the logarithmic behavior.

It is interesting to contrast the $\sim-\ln{x}$ to $\sim{x}^{-1}$ crossover with the isotropic Kondo
model, in which case a crossover from $\sim{x}^{-1}$ to $\sim{x}^{-2}$ has been reported for
spin-spin correlators \cite{Borda}.  
Our results for the RLM also provide definite predictions for the screening cloud of the
single-impurity Anderson model, when described by two-point impurity-bath correlators.  Some
available numerical data is consistent with this prediction, but a thorough exploration of the SIAM
screening cloud using various correlators is clearly necessary.

In addition to the standard RLM, we have presented results for the screening cloud with more
complicated Fermi surface topology, as can be realized with next-nearest-neighbor couplings in the
tight-binding bath.  An intriguing case arises when part of the Fermi surface disallows
linearization because the dispersion curves touches the chemical potential without crossing it; the
divergent density of states leads to a non-Fermi-liquid behavior, namely, a power-law
($\sim{x}^{-3/2}$) decay at large distances with exponent $\neq1$.  It is remarkable that such
unconventional physics can be generated from free-fermion systems.

\ack


MH thanks I.~Affleck, E.~Boulat and M.~Garst for useful discussions.  The authors also thank F.~Assaad and
F.~Goth for help with the data of Ref.~\cite{Goth_Assaad}.

\appendix

\section{\label{sec_appendix_greenfuncion_analytic} One-particle Green's function derivation \label{sec:One-particle-Green's-function}}

We present here the explicit analytical calculation for external RLM leading to Eq.(\ref{gdc3}).  
Similar results can be obtained for the other two geometries.  We also show that Eq.(\ref{gdc1})
is valid even in the presence of interactions at the impurity site.

At finite temperature, the imaginary-time single-particle Green's
function are given by 
\begin{eqnarray}
G(\tau) & = & \left(\begin{array}{cc}
G_{cc}(\tau) & G_{cd}(\tau)\\
G_{dc}(\tau) & G_{dd}(\tau)
\end{array}\right)
\end{eqnarray}
with $\left\langle ...\right\rangle $ the thermal average with respect
to the Gibbs ensemble with Hamiltonian $H_{ext}$ and where $G_{cc}\left(\tau\right)$,
$G_{cd}\left(\tau\right)$, $G_{dc}\left(\tau\right)$ and $G_{dd}\left(\tau\right)$
(a $\mathcal{L}\times \mathcal{L}$ matrix, a column vector, a line vector and a c-number
respectively) are defined by 
\begin{eqnarray}
\left[G_{cc}\left(\tau\right)\right]_{i,j} & = & \left\langle T_{\tau}c_{i}(\tau)c_{j}^{\dagger}(0)\right\rangle \\
\left[G_{cd}\left(\tau\right)\right]_{i} & = & \left\langle T_{\tau}c_{i}(\tau)d^{\dagger}(0)\right\rangle \\
\left[G_{dc}\left(\tau\right)\right]_{i} & = & \left\langle T_{\tau}d(\tau)c_{i}^{\dagger}(0)\right\rangle \\
G_{dd}(\tau) & = & \left\langle T_{\tau}d(\tau)d^{\dagger}(0)\right\rangle .
\end{eqnarray}
These definitions extend to the other two geometries, and also to the
interacting cases.  (For the Anderson model the additional spin
index has to be taken into account, however off diagonal correlations
in spin index vanish identically by spin conservation). 

Assuming that interactions arise only at the impurity site, Dyson's
equation in Matsubara space, takes the form 
\begin{eqnarray}
G_{dd}^{-1}(i\omega_{n}) & = & G_{0,dd}^{-1}(i\omega_{n})-\Sigma_{dd}\left(i\omega_{n}\right)\label{eq:gfr-1}\\
G_{dc}(i\omega_{n}) & = & -G_{dd}(i\omega_{n})VG_{0,cc}(i\omega_{n})\label{gfr}\\
G_{cc}(i\omega_{n}) & = & G_{0,cc}(i\omega_{n})+\\
 &  & G_{0,cc}(i\omega_{n})V^{\dagger}G_{dd}(i\omega_{n})VG_{0,cc}(i\omega_{n})\nonumber 
\end{eqnarray}
and thus the problem is reduced to finding the explicit form of $G_{dd}(i\omega_{n})$.
Here $G_{0,cc}$ and $G_{0,dd}$ are the single-particle Green's function
of the conduction electron bath and impurity site respectively in
absence of tunneling between the two (\textit{i.e.} $J'=0$) 
\begin{eqnarray}
\left[G_{0,cc}(i\omega_{n})\right]_{i,j} & = & \sum_{k}\frac{e^{ik(r_{i}-r_{j})}}{i\omega_{n}-(\epsilon_{k}-\mu)}\\
G_{0,dd}(i\omega_{n}) & = & \frac{1}{i\omega_{n}-(\epsilon_{d}-\mu)}
\end{eqnarray}
with $\epsilon_{d}$ is the energy of the impurity level and $\mu$
is the chemical potential of the bath, and  $V$ is a line vector
with entries $\left[V\right]_{i}=-J'\delta_{i,0}.$ 
%

Eq.(\ref{gdc1}) can be obtained form Eq.(\ref{gfr}) noting that
$\left\langle d^{\dagger}c_{i}\right\rangle =-G_{dc}(\tau=0^{-})$
and by using the spectral decomposition of the impurity Green's function
$G_{dd}(i\omega_{n})=\int d\nu\frac{A_{dd}(\nu)}{i\omega_{n}-\nu}$, 
where $A_{dd}(\nu)=-1/\pi\mathrm{Im}\left[G_{dd}\left(\omega+i0^{+}\right)\right]$
is the impurity spectral function. 

For the external RLM the self-energy of the $d$ electrons acquires
the simple form 
\begin{eqnarray}
\Sigma_{dd}\left(i\omega_{n}\right) & = & VG_{0,cc}(i\omega_{n})V^{\dagger}
\end{eqnarray}
In the large $\mathcal{L}$ limit using the approximation $\frac{1}{\mathcal{L}}\sum_{k}...\to\int
d\nu\,\mbox{\ensuremath{\rho\left(\nu\right)}}...$ and assuming a constant density of states
$\rho\left(\nu\right)\simeq\rho_0\Theta\left(\Lambda-\left|\nu-\mu\right|\right)$
around the Fermi level, one obtains, in the wide band limit $\nu\ll\Lambda$,
$\Sigma_{dd}\left(i\omega_{m}\right)\simeq-i\mathrm{sign}\left(\omega_{n}\right)\Gamma$ with
$\Gamma=\pi J'^{2}\rho_0$. This result toghether with Eq.(\ref{eq:gfr-1}) yields the expression
of the spectral function of the $d$ level used in the main text.

In the case of Fig.\ref{FigNNN}, where the dispersion relation touches the Fermi energy without crossing it,
 the impurity self-energy, for energies much smaller than the bandwidth,  is given by:
 \begin{eqnarray}
 \Sigma_{dd}(i\omega_n) &=& -i~\mathrm{sign}\left(\omega_{n}\right)\Gamma+\frac{\delta}{\sqrt{i\omega_n}},  \non 
\end{eqnarray}
with $\delta=\frac{J'^2}{2\sqrt{a_F}}$ and where $a_F$ is obtained by expanding the dispersion relation about 
the critical Fermi momentum (here assumed to arise for $k=0$)   $|\epsilon_k-\mu| \simeq a_F k^2$ .   
$\Gamma = \pi J'^2 \rho_0$ is given as before with $\rho_0 = \left|\pi v_F\right|^{-1}$ being the regular part of the density of states. 
The corresponding impurity spectral function is given by:
%

\begin{eqnarray}
A_{dd}(\nu) =  \cases{
 \frac{1}{\pi}\frac{\Gamma}{(\nu +\frac{\delta}{\sqrt{\nu}})^2+\Gamma^2} & for $\nu>0$  \\
\frac{1}{\pi}\frac{\Gamma+\frac{\delta}{\sqrt{\nu}}}{\nu^2+(\Gamma+\frac{\delta}{\sqrt{\nu}})^2} &  for $\nu<0$
}
\end{eqnarray}

\end{document}